%% file: main.tex
\documentclass[runningheads]{llncs}
\usepackage[T1]{fontenc}
\usepackage[caption=false]{subfig}

\usepackage{graphicx}
\usepackage{hyperref}
\usepackage{color}

\usepackage[table,xcdraw]{xcolor}

\begin{document}
\title{A New Framework for Error Analysis in Computational Paleographic Dating of Greek Papyri}
\titlerunning{A New Framework for Error Analysis}
%
\author{
Giuseppe De Gregorio\inst{1}\orcidID{0000-0002-8195-4118} \and
Lavinia Ferretti\inst{1}\orcidID{0009-0005-2619-117X} \and
Rodrigo C. G. Pena\inst{1}\orcidID{0000-0002-9010-2830} \and
Isabelle Marthot-Santaniello\inst{1}\orcidID{0000-0003-0407-8748} \and
Maria Konstantinidou\inst{2}\orcidID{0000-0002-8744-1444} \and
John Pavlopoulos\inst{3}\orcidID{0000-0001-9188-7425}
}
\authorrunning{G. De Gregorio et al.}
%
\institute{
University of Basel, Switzerland, \email{\{giuseppe.degregorio,lavinia.ferretti, rodrigo.cerqueiragonzalezpena,i.marthot-santaniello\}@unibas.ch}\and
Democritus University of Thrace, Greece,
\email{mkonst@helit.duth.gr} \\ \and
Athens University of Economics and Business, Greece, \email{annis.pavlo@aueb.gr}
}
\maketitle              
%
%
%
\input{texs/00_abstract}
\section{Introduction}
\input{texs/01_Introduction}
\section{Related Works}
\input{texs/02_RelatedWorks}

\section{The Methodology}
\input{texs/03_TheMethodology}
\section{Results}
\input{texs/04_Results}
\section{Discussion}
\input{texs/07_Discussion}
\section{Conclusion}
\input{texs/08_Conclusion}

%
%
%
\bibliographystyle{splncs04}
\bibliography{ref}

\end{document}

%% file: texs/00_abstract.tex
\begin{abstract}

The study of Greek papyri from ancient Egypt is fundamental for understanding Graeco-Roman Antiquity, offering insights into various aspects of ancient culture and textual production. Palaeography, traditionally used for dating these manuscripts, relies on identifying chronologically relevant features in handwriting styles yet lacks a unified methodology, resulting in subjective interpretations and inconsistencies among experts. 
Recent advances in digital palaeography, which leverage artificial intelligence (AI) algorithms, have introduced new avenues for dating ancient documents. This paper presents a comparative analysis between an AI-based computational dating model and human expert palaeographers, using a novel dataset named Hell-Date comprising securely fine-grained dated Greek papyri from the Hellenistic period. 
The methodology involves training a convolutional neural network on visual inputs from Hell-Date to predict precise dates of papyri. In addition, experts provide palaeographic dating for comparison. To compare, we developed a new framework for error analysis that reflects the inherent imprecision of the palaeographic dating method. The results indicate that the computational model achieves performance comparable to that of human experts. These elements will help assess on a more solid basis future developments of computational algorithms to date Greek papyri.

\keywords{Greek papyri  \and Computational dating \and Palaeography \and Error analysis \and Human comparison.}
\end{abstract}

%% file: texs/01_Introduction.tex
Greek papyri preserved thanks to the dry climate of Egypt represent an unparalleled primary source for the study of Graeco-Roman Antiquity. These texts, which cover a wide range of contents from documentary testimonies (contracts, letters...) to literary works, play a crucial role in our understanding of ancient culture in general and of handwriting evolution and book production in particular. A rough classification of papyri divides them between documentary and literary texts: while the former are usually written in fast, informal cursive scripts, the latter use formal scripts, i.e. slow, detached and easy-to-read, sometimes calligraphic scripts which received the designation of "book hands". Regardless of the type of papyrus, the informational value of such manuscripts significantly increases when their dating can be established. Sometimes, it is possible to assign them a date thanks to textual or archaeological evidence; when such clues are lacking, researchers resort to palaeography to estimate an approximate and broad date for their production.

Palaeographers operate under the assumption that writing styles share some features among coeval specimens and gradually change over time. For dating, scholars use a comparative technique: a date is proposed by comparison with other previously dated samples, preferably but not necessarily with objectively dateable papyri. Although multiple palaeographers identify some specific writing features as chronologically significant, to date, no unified methodology is consensual among experts. This results in a situation where each palaeographer is free to focus on aspects of writing they deem most significant. Moreover, even when common features can be identified, they are rarely objectively measurable or calculable~\cite{cavallo_2008,cavallo_2009,harrauer_2010,orsini_clarysse_2012,nongbri_2019}. This palaeographical method, heavily relying on personal expertise that can be acquired from a long acquaintance with manuscripts, is open to many uncertainties, if not errors, and is indisputably difficult to communicate. This explains the regular occurrence of conflicting results among different experts \cite{harrauer_2010,orsini_clarysse_2012,choat_2019}.

Recently, digital palaeography has introduced sophisticated techniques, including image analysis using artificial intelligence (AI) algorithms. However, interpreting the results obtained from such techniques requires caution, as the features used for making a decision in trained models may differ from those relevant to human experts.

In this work, we aim to provide a dating framework for Greek papyri that incorporates the inherent imprecision of chronological attribution only based on handwriting. To this end, we have:

\begin{enumerate}
    \item compiled a new dataset, named Hell-Date, composed of images of papyri whose exact year of writing is established thanks to unequivocal internal evidence;

    \item evaluated the performance of a convolutional neural network trained on visual inputs, specialised in dating ancient documents when applied to a dataset characterised by precise and granular dating (Hell-Date);
    
    \item set an experiment based on the network pipeline to evaluate the results of the model compared with that of expert scholars; 
    
    \item adopted metrics that integrate the chronological imprecision inherent to the method to analyse the error.

\end{enumerate}

%% file: texs/02_RelatedWorks.tex
Various techniques have been applied in the past to the challenge of dating ancient documents using computational methods, changing according to factors such as document type, language, and historical period.

In the literature, diverse standard Machine Learning techniques have been tried. Dhali et al.~\cite{DHALI2020413} used a Support Vector Regressor-based technique to date a collection of 595 Dead Sea Scrolls written in the Hebrew alphabet, ranging from 250 to 135 BCE, through feature extraction from manuscript scripts. Adam et al.~\cite{adam2018kertas} proposed a sparse representation-based method to date historical handwritten Arabic manuscripts, employing a K-nearest neighbour (KNN) approach. Some methodologies rely on textual analysis for dating, as Baledent et al.~\cite{baledent-etal-2020-dating}, who introduced a dataset that features numerous ancient documents in French and used decision trees and random forests at both character and token levels.

Noteworthy advancements have emerged from techniques rooted in convolutional networks and Deep Learning models. Li et al.~\cite{li2015publication} presented an approach using convolutional neural networks (CNNs) alongside text features extracted via optical character recognition for estimating the publication date of historical English printed documents from the 15\textsuperscript{th} to the 19\textsuperscript{th} century. 
Cloppet et al.~\cite{Cloppet2017} and Seuret et al.~\cite{Seuret2021} propose competitions whose tasks include dating manuscripts and printed material from the medieval ages. Hamid et al.~\cite{hamid2019deep} leveraged pre-trained CNNs to date images of medieval Dutch cards from the 14\textsuperscript{th} to the 16\textsuperscript{th} century, with a focus on image fragments. Wahlberg et al.~\cite{Wahlberg2016hist} proposed a deep learning methodology for dating pre-modern handwritten documents, achieving results comparable to human experts on a dataset comprising over 10,000 medieval Swedish cards.

While various methodologies have been explored across different languages and historical periods, fewer studies have focused on texts written on papyrus in ancient Greek. Pavlopoulos et al.~\cite{pavlopoulos-etal-2023-dating} experimented on dating Greek papyri analysing textual contents using regression methodologies. One of the first approaches to dating Greek papyrus images via convolutional networks was taken by Paparrigopoulou et al.~\cite{paparrigopoulou2023greek}, reporting an average dating error of more than a century. Subsequent work by Pavlopoulos et al.~\cite{pavlopoulos2023explaining}, focusing on literary papyri from the Roman period (1\textsuperscript{st} - 4\textsuperscript{th} CE), demonstrated improved accuracy through a segmentation strategy. The authors showed that segmenting to line level rather than working at the whole document level reduces the average dating error.

%% file: texs/03_TheMethodology.tex
\subsection{Datasets}
\input{texs/03.1_data}

\subsection{The Computational Dating}
\input{texs/03.2_AI}

\subsection{The Human Experts Dating}
\input{texs/03.4_Humans}

%% file: texs/03.1_data.tex
\subsubsection{The New Hell-Date Dataset}
\begin{figure}[ht]
    \centering
    \includegraphics[width=1\linewidth]{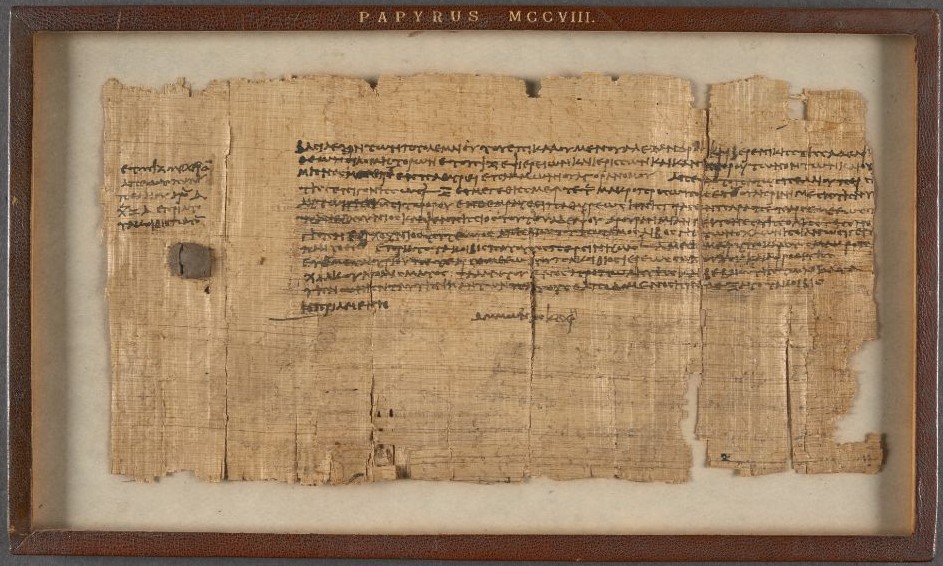}
    \caption{Image of the papyrus TM~244 $=$ P.Lond. III 1208 $=$ P.Lond. inv. 1208. Image courtesy of the British Library. The first 3 lines of column 2 translate: "\textit{When Ptolemy, also known as Alexander, and Berenice, his sister, the mother-loving gods, are reigning for the 17\textsuperscript{th} year, under the priests and priestesses and the canephoros which are in charge, day 4 of the month Mecheir, in Pathyris, under the notary Ammonios}".}
    \label{fig:pap_example}
\end{figure}

Hell-Date is a dataset composed of 187 images of 155 papyri written in Greek and dated to the Hellenistic period (from the late 4\textsuperscript{th} to the 1\textsuperscript{st} c. BCE)\footnote{The dataset is accessible at the following link:~\href{https://d-scribes.philhist.unibas.ch/en/hell-date/}{https://d-scribes.philhist.unibas.ch/en/hell-date/}.}. These texts are securely dated, i.e. they contain textual evidence that points to the exact year in which they were written - usually the mention of the date somewhere in the text. For instance, the papyrus TM~244 shown in Figure~\ref{fig:pap_example} is a contract in which the first 3 lines of column 2 explicitly mention that it was written in the 17\textsuperscript{th} year of reign of King Ptolemy Alexander and Queen Berenice, the fourth day of the month Mecheir~\footnote{Papyri are cited according to Trismegistos (TM) Numbers, for which cf.~\cite{trismegistos} and \href{https://www.trismegistos.org/about_how_to_cite.php}{$https://www.trismegistos.org/about\_how\_to\_cite.php$}.}. This date, converted into the modern system, corresponds to February 17, 97 BCE.

Several dating systems are used in Hellenistic papyri, and none, as expected, strictly aligns with our modern Gregorian calendar~\cite{bagnall_2009,Bennet_2011,Paulissen_Vandorpe_2019}. It is thus impossible to give one Gregorian year of writing for some papyri; for instance, papyrus TM~121853 was written in the month of Audnaios of the 18\textsuperscript{th} year of reign of king Ptolemy II, which corresponds to a period from late December 268 to late January 267 BCE. In these cases, less than twenty in the full dataset, we have given both years in the dataset metadata but used arbitrarily the oldest year for the purposes of this paper.

Restricted to the Hellenistic period (from~-310 to~-3), Hell-Date was selected on purpose to have an even distribution with about 50 documents per century. These texts are written in so-called "documentary hands", i.e. fast scripts that prioritise speed over legibility and aesthetics. Such scripts are often cursive, present ligatures and can be hard to read.

The images incorporated within the dataset originate from diverse online collections and resources undergoing digitisation over a span exceeding two decades, using distinct imaging protocols that are not documented\footnote{For a list of these resources, see the online description of the dataset: \href{https://d-scribes.philhist.unibas.ch/en/hell-date/dataset/}{https://d-scribes.philhist.unibas.ch/en/hell-date/dataset/}.}. Consequently, they exhibit considerable property variations, notably about scaling to actual size, colour capture, resolution, and bit depth. Moreover, due to the condition of the original papyrus, these images may exhibit noise, including surface damage, empty spaces, and delineations marking the papyrus edges. An example of such problems can be seen in Figure~\ref{fig:pap_example}, where there are stains and holes on the surface, empty spaces between columns and a modern frame with modern writing. All the collected images were segmented into lines of text using the docExtractor segmentation tool~\cite{docextractor} followed by a manual correction. Subsequently, lines longer than twice the average length were further divided, while short lines that contained only a few characters or single words were discarded. This process resulted in clippings of text lines with comparable dimensions, without any layout information such as column width or document size, and yielded a total of 8,230 images of text lines, with a peak of distribution in the late 2\textsuperscript{nd} c. BCE.

\subsubsection{Comparable dataset}
For comparison, we selected from the existing literature the Papyri Literary Lines (PLL) dataset \cite{pavlopoulos2023explaining}. To the best of our knowledge, this is the only existing dataset tailored for the task of dating Greek papyri, in this case, from the Roman period. It is composed of 2,774 line images extracted from 159 images of Greek papyri, selected from the~\href{https://classics.artsandsciences.baylor.edu/academics/greek-bookhands-database}{\textit{Collaborative Database of Dateable Greek Bookhands}}; for details on the PLL dataset, one could refer to the original publication.

The PLL dataset is different from Hell-Date. These papyri date between the 1\textsuperscript{st} and the 4\textsuperscript{th} c. CE. Their date can only be estimated with some approximation; therefore, it is only given at a century level. The chronological distribution is uneven, with a peak in the 2\textsuperscript{nd} and 3\textsuperscript{rd} c. CE. They are written in book hands, i.e. formal scripts usually used for literary texts. These hands may present artificial phenomena like archaism or fossilisation that can complicate their dating. Last, the PLL dataset contains almost three times fewer line images than Hell-Date. The characteristics of the two datasets are summarised in Table~\ref{tab_datasets}.

For the experiments, the datasets were partitioned such that $90\%$ of the document images were allocated to the training set, while the remaining $10\%$ were designated for the test set. Importantly, this division was performed at the level of document image rather than the text line. This approach ensures that no text line present in the test set originates from a document image used for training. Following the construction of the test set, a validation set comprising $10\%$ of the training data is extracted.  
At the end, the Hell-Date training set comprises 7424 line images from 169 documents and the test set of 806 lines from 18 documents. Conversely, the PLL training set is composed of 2496 line images from 146 documents, and the test set of 270 lines from 13 documents.

\begin{table}[ht]
\centering
\caption{Datasets overview.}
\label{tab_datasets}
\begin{tabular}{lcc}
\hline
      & \textbf{Hell-Date} & \textbf{PLL}  \\ 
\hline\hline       

\textit{Period} & Hellenistic & Roman \\
\textit{Centuries} & 4\textsuperscript{th}-1\textsuperscript{st} BCE & 1\textsuperscript{st}-4\textsuperscript{th} CE \\
\textit{Precision of Date} & to the Year &  to the Century\\
\textit{Type of Writing} & Cursive Hand  & Book Hand \\

\textit{N Documents} &  187   &  159    \\
\textit{N  Lines }   & 8230   &  2774    \\                

\hline
\end{tabular}
\end{table}

%% file: texs/03.2_AI.tex
\subsubsection{The Model}
\input{texs/03.2.1_Model}

\subsubsection{}
 Bearing in mind that the Hell-Date dataset provides a more granular ground truth than PLL, we
 experimented with several scenarios with various combinations of Hell-Date and PLL to evaluate the relevance of transfer learning on the performance of computational dating. 
  
 First, we experimented with training and testing the network on Hell-Date alone and on PLL alone. Second, initial training uses one dataset, and further training uses the other one, testing on the last one. This means, in the first case we train first on PLL and then we continue on Hell-Date and test on Hell-Date, and in the second case, we train on Hell-Date, then on PLL and test on PLL. Finally, we first merged the Hell-Date and the PLL training sets and then, we trained the network on this combination. We tested on both test sets and on the combination of test sets.
 As the PLL and Hell-Date datasets differ in size, we explored the partial use of the larger dataset to maintain data balance in the union dataset.

%% file: texs/03.2.1_Model.tex
The convolutional network used in this study is referred to as fCNN, originally introduced by Pavlopoulos et al.~\cite{pavlopoulos2023explaining}\footnote{The code published in the article is available at \href{https://github.com/ipavlopoulos/palit}{https://github.com/ipavlopoulos/palit}.}. The fCNN network is presented in two versions: fCNNc as a classifier and fCNNr as a regressor. Employing the model as a classifier requires a rigid discretisation of the time axis to define distinct classes. However, the discretisation of time implies the loss of granular information on exact dates. This conflicts with the primary aim of achieving the most precise dating feasible, particularly considering the granularity of chronological ground truth in the Hell-Date dataset. Therefore, the fCNNr regressor version aligns well with our goals.

The model architecture consists of two Conv2D layers featuring 32 and 64 channels respectively, followed by a 3-layer feed-forward neural network culminating in a single output neuron responsible for date estimation. Convolutional operations are defined by a kernel size of 5, stride of 1, zero padding, and subsequent maximum pooling with a 2x2 window. The feed-forward component processes a flattened representation obtained from the convolutional layers, sequentially reducing the neuron count to 1024 and then 512 before the final date prediction. Each layer incorporates Rectified Linear Unit (ReLU) activation functions. Functioning as a regression model, this network takes as input the image of a handwritten text line and produces a single numerical output representing the inferred date.

During the model training phase, data augmentation techniques are employed to enhance network robustness. This involved random deletion of image fragments with a probability of 0.5, replacing the pixel values with 0.5. Additionally, images undergo transformations including Gaussian blur with a kernel size of 3 and random affine transformations up to 3\textsuperscript{rd} degree. Furthermore, random cropping and resizing are applied to each image while maintaining a 1:6 aspect ratio to minimise excessive alteration of the manuscript content within the images.

%% file: texs/03.4_Humans.tex
Expert scholars in the study of ancient manuscripts can ascertain the approximate age of documents based on the palaeographic analysis of handwritten text fragments. In this study, we recruited five highly experienced scholars specialising in Hellenistic Greek papyri 
to participate in a dating assessment.

To arrange the experiment, we selected three lines from each document within the Hell-Date test set. This number allows, on the one hand, reflecting the variety and average preservation of the original document and, on the other hand, not requiring too much time from the participants. We carefully selected lines that do not contain any textual clue on the date of the text. We created a form presenting the images in random order, asking the respondents to provide for each line a dating interval expressed as two integers (starting and ending years of the interval). The respondents were asked to be as specific in their dating as they felt confident and to base themselves only on the appearance of the writing. They only had one chance to fill out the form and were given thirty minutes to answer, so that they could not try to identify in the literature the text from which the line came. They were not asked to justify their dating. Although dating based on lines is uncommon in traditional palaeography, the experiment was devised as such to allow for comparability with the results of the computational model.

%% file: texs/04_Results.tex
In this section, we present the results of our experiments, beginning with the results obtained through the computational dating method. Subsequently, we provide the results obtained by the human experts.

\subsection{Computational Dating}
For training the various models, we employed Adam optimisation with a learning rate of 1e-3 and a batch size of 16. Training iterations extended up to 200 epochs, implementing an early stopping policy with patience of 20 epochs. The loss function employed throughout the training was the mean squared error. Regarding representation, floating point numbers are chosen to represent centuries, such that the integer part represents the exact year (e.g.  1.50 corresponds to the year 150). Consequently, Mean Absolute Error (MAE) and Mean Squared Error (MSE) are adopted as the evaluation metric for error assessment.

Table~\ref{tab:results_normal_training} presents the Mean Absolute Error (MAE) and the Mean Squared Error (MSE) when the network is trained and tested on the same dataset. Given the larger dimension of the Hell-Date dataset compared to PLL, training experiments were repeated with Hell-Date while considering different percentages of the training set each time.
As can be observed from the error reported in the table, the network achieves a MAE of approximately 55 years for both datasets. However, these outcomes are attained solely when using $100\%$ of the available data for training. Indeed, when the size of the training data between PLL and Hell-Date is comparable (i.e. when approximately $35\%$ of the Hell-Date training set is considered), a larger error is noted on Hell-Date.

\begin{table}[ht]
\centering
\caption{Number of text line images for training, validation, test and results in terms of MAE and MSE.}
\label{tab:results_normal_training}
\begin{tabular}{lccccc}
\hline
                        & \textbf{PLL}   & \textbf{Hell-Date} & \textbf{Hell-Date} & \textbf{Hell-Date} & \textbf{Hell-Date}  \\
                        & \textbf{100\%} & \textbf{100\%} & \textbf{50\%} & \textbf{35\%} & \textbf{10\%}  \\ 
\hline\hline
\textit{Trainings Lines}    &  2496 &  7424 &  3340 &  2338 &  667  \\
\textit{Validation Lines}   &  270  &  806  &  372  &  260  &  75  \\    
\textit{Test Lines}         &  270  &  806  &  806  &  806  &  806  \\
\hline
\textit{MAE}                &  \textbf{0.5418}  &  \textbf{0.5637}  &  0.5923 &  0.6050 &  0.6446  \\
\textit{MSE}                &  \textbf{0.5161}  &  \textbf{0.4922 } &  0.5274 &  0.5776 &  0.5965  \\      
\hline
\end{tabular}
\end{table}

\subsubsection{Can Transfer Learning Improve Performances?}
At this point, we proceed to assess the feasibility of transferring learning between the two datasets. 
In this context, our approach involves initially training the network on one of the two distinct datasets before proceeding with additional training on the other dataset. Table~\ref{tab:results_pre_training_HD} displays the results achieved on the Hell-Date test set after training conducted initially on PLL followed by training on Hell-Date. Table~\ref{tab:results_pre_training_PLL} exhibits the results on the PLL test set when the network undergoes training initially with Hell-Date followed by training with PLL.
Comparing the last tables with the results in Table~\ref{tab:results_normal_training}, it becomes evident that the transfer of learning yielded marginal improvements in dating the Hell-Date test set, leading to a slight decrease in both the MAE and the MSE. Conversely, there was a decline in performance when evaluating the PLL documents.

\begin{table}[ht]
\centering
\caption{Results in terms of MAE and MSE on Hell-Date test set when the network is trained on PLL and further trained on a percentage of Hell-Date.}
\label{tab:results_pre_training_HD}
\begin{tabular}{lccccc}
\hline
   & \textbf{Hell-Date}   & \textbf{Hell-Date} & \textbf{Hell-Date} & \textbf{Hell-Date} & \textbf{Hell-Date}  \\ 
   & \textbf{0\%} & \textbf{25\%} & \textbf{50\%} & \textbf{75\%} & \textbf{100\%}  \\ 
\hline\hline
\textit{MAE}   &  9.9426  &  0.6452  &  0.6390 &  \textbf{0.5357} &  0.5424  \\
\textit{MSE}   &  133.88  &  0.6568  &  0.6669 &  0.6669 &  \textbf{0.4883}  \\      
\hline
\end{tabular}
\end{table}

\begin{table}[ht]
\centering
\caption{Results in terms of MAE and MSE on PLL test set when the network is trained on Hell-Date and further trained on a percentage of PLL.}
\label{tab:results_pre_training_PLL}
\begin{tabular}{lccccc}
\hline
   & \textbf{PLL}   & \textbf{PLL} & \textbf{PLL} & \textbf{PLL} & \textbf{PLL}  \\ 
   & \textbf{0\%} & \textbf{25\%} & \textbf{50\%} & \textbf{75\%} & \textbf{100\%}  \\ 
\hline\hline
\textit{MAE}   &  96.089  &  0.6505  &  0.5904 &  0.5857 &  \textbf{0.5687 } \\
\textit{MSE}   &  51342   &  0.6315  &  0.5689 &  0.5664 &  \textbf{0.5362}  \\ 
\hline
\end{tabular}
\end{table}

Finally, we attempt to train the network by combining the two datasets. In this experiment, training is conducted on a set comprising the union of the training sets from the different datasets. Subsequently, the trained network is assessed on the individual test sets of the two datasets as well as on the combined test set.
Given the considerable size discrepancy between the Hell-Date and PLL test sets, we iterate the training process while adjusting the percentage of Hell-Date used for fusion; the two training sets are fairly balanced when considering approximately $35\%$ of the Hell-Date training set. The outcomes of this experiment are summarised in Table~\ref{tab:results_joinDS}.

\begin{table}[ht]
\centering
\caption{Results in terms of MAE and MSE when the network is trained on the union between PLL and Hell-Date}
\label{tab:results_joinDS}
\begin{tabular}{cc|cc|cc|cc}
\hline 
    \textbf{PLL}& \textbf{Hell-Date}& \multicolumn{2}{c|}{\textbf{Test Set}}& \multicolumn{2}{c|}{\textbf{Test Set}} & \multicolumn{2}{c}{\textbf{Test Set}} \\

    \textbf{\%}&\textbf{\%}& \multicolumn{2}{c|}{\textbf{PLL}}& \multicolumn{2}{c|}{\textbf{Hell-Date}} & \multicolumn{2}{c}{\textbf{PLL + Hell-Date}} \\
    \hline\hline
   &&\textbf{MAE}&\textbf{MSE}&\textbf{MAE}&\textbf{MSE}&\textbf{MAE}&\textbf{MSE}\\
\cline{3-8} 
100&100&1.0612&2.2559&\textbf{0.7918}&\textbf{1.4973}&\textbf{0.8756}&\textbf{1.7332}\\
100&50 &0.8249&1.3885&0.9097&1.9220&0.8834&1.7560\\
\rowcolor[HTML]{D0D0D0} 
100&35 &0.9186&1.6376&0.9010&1.7588&0.9095&1.7211\\
100&10 &\textbf{0.7568}&\textbf{0.8545}&1.2530&3.3185&1.0987&2.5512\\
\hline
\end{tabular}
\end{table}

As depicted in Table~\ref{tab:results_joinDS}, the percentage used for Hell-Date significantly impacts the results. The optimal results on the combined test sets are achieved when employing $100\%$ of both training sets, on the other hand, it is evident that the error distribution is uneven across the two datasets. Notably, when the two training sets are balanced in terms of size (approximately $35\%$ of Hell-Date), the errors on the two test sets are close. By comparing the results of Table~\ref{tab:results_normal_training}, \ref{tab:results_pre_training_HD}, \ref{tab:results_pre_training_PLL}, and \ref{tab:results_joinDS},
it appears that merging the datasets does not bring any improvement: the results are even the worst we obtained in all our experiments.  The different nature of the handwriting in the two datasets may explain this behaviour.

\subsection{The Human Experts Dating}
As explained above, five expert scholars were provided with a form to date 54 line images (3 lines for each of the 18 images of the Hell-Date test set)\footnote{The values for individual respondents are accessible at the following link: \href{https://d-scribes.philhist.unibas.ch/en/hell-date/}{https://d-scribes.philhist.unibas.ch/en/hell-date/}.}. Some experts opted not to provide answers for specific images presented in the form. To compute the MAE, for these unanswered entries, the error value corresponding to the maximum error within the dataset was used as an answer.
Table~\ref{tab_Human_MAE} presents the MAE observed in the responses of each expert, along with the average error calculated from the collective performances. Additionally, the table shows the error calculated by disregarding the rows where the expert did not submit a response.

\begin{table}[ht]
\centering
\caption{MAE obtained from the dating of the expert scholars, by including and excluding empty answers.}
\label{tab_Human_MAE}
\begin{tabular}{p{4cm}ccccc}
\hline 
 \bf Expert& \textbf{01} & \textbf{02} & \textbf{03} & \textbf{04} & \textbf{05}\\ 
\hline\hline
\textit{MAE} & 1.27 & 0.41 & 2.28 & 0.48 & 0.62 \\
\textit{MAE (excl. empty answers)} & 1.13 & 0.41 & 1.75 & 0.48 & 0.53 \\
\hline
\textit{\# empty answers} & 3 & 0 & 20 & 0 & 2 \\
\hline
\end{tabular}
\end{table}

Given the variability in errors among different experts, one may wonder about the extent of agreement across these experts. To address this inquiry, Table~\ref{tab:AII} presents some indices to measure the level of agreement between experts: the Mean Pairwise MAE, the Mean Pairwise Spearman and Pearson Correlation, and the Fleiss' kappa index. The first is computed by calculating the MAE considering the dates of one expert as truth and those of the others as predictions. The second and third compute the correlations between any two experts and average them. Finally, to compute the kappa index, it was necessary to discretise the responses to enable the assignment of shared labels. To achieve this, the time axis was discretised with a step size of 25 years. Subsequently, for each document, a positive (1) or negative (0) label was assigned to each interval of the discretised time axis based on the expert's prediction, such that intervals covered partially or totally by the prediction were set to 1, and intervals fully outside the prediction to 0. 
As can be seen from the table, we can record a very slight agreement among the experts, being the Mean Pairwise MAE high, the Mean Pairwise correlations indices close to 0.5, and Fleiss' Kappa slightly greater than zero.

\begin{table}[ht]
\centering
\caption{Indices to measure the agreement among experts.}
\label{tab:AII}
\begin{tabular}{p{5cm}cc}
\hline 
 \bf Index& \textbf{Value} \\
\hline\hline
\textit{Mean Pairwise MAE}           &  105.91  \\
\textit{Mean Pairwise Spearman Corr} &    0.54  \\
\textit{Mean Pairwise Pearson Corr}  &    0.53  \\
\textit{Fleiss' Kappa }              &    0.03  \\
\hline
\end{tabular}
\end{table}

%% file: texs/07_Discussion.tex
\subsection{Error Analysis}

In the context of automatic dating of historical documents, error is a significant challenge to address. Analysis based on Mean Absolute Error may be limited, as a writing style does not correspond to a single precise date, but rather to a time interval that reflects the gradual process of handwriting evolution. To address this challenge, we employed the method of the Error Time Window (ETW). The ETW is defined as a rectangular window function $\mathrm{\Pi}$, centred around the production year of the document $Y$, characterised by a certain width $\alpha$ in terms of time $t$ expressed in years:
\begin{equation}
ETW = \mathrm{\Pi}(\alpha t-Y)    
\end{equation}
Using this window, we evaluate whether a prediction obtained by the neural network falls inside or outside this range. 

\begin{figure}[ht]
\centering
\includegraphics[width=0.70\textwidth]{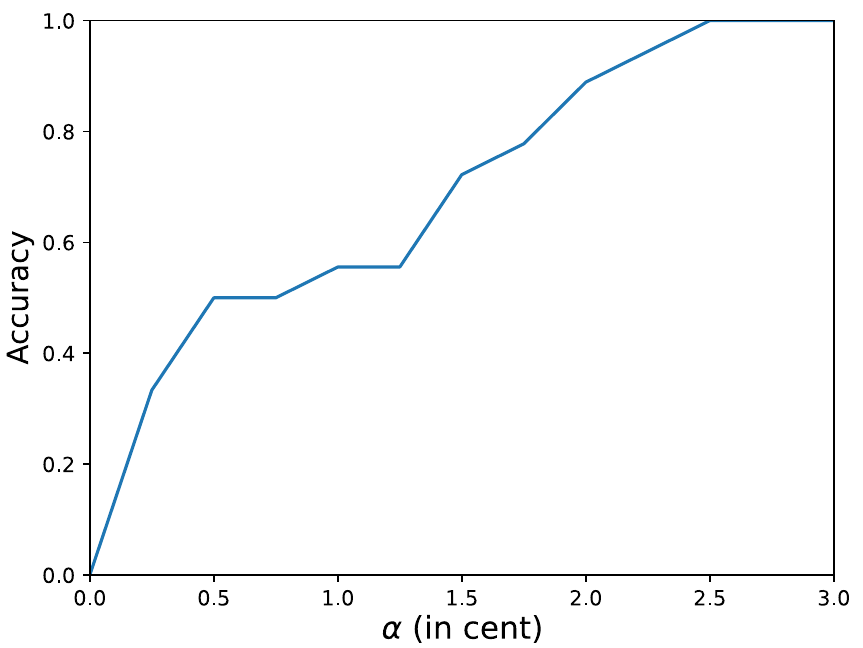}
\caption{ Accuracy of the Hell-Date test set dating according to the size $\alpha$ of the ETW.}
\label{fig:error_wind}
\end{figure}

This approach enables the calculation of the Accuracy index ($A(\alpha)$), representing the proportion of correct predictions compared to the total predictions made for a given $\alpha$ width of the ETW:
\begin{equation}
\label{eq:prevalence}
A(\alpha) = \frac{P}{P+N}
\end{equation}
Here, $P$ (positive) represents correct predictions for each line, and $N$ (negative) indicates incorrect predictions relative to the defined time window.
Figure~\ref{fig:error_wind} illustrates the accuracy trend as the size $\alpha$ of the ETW varies, considering the model that performed best on the dating of the Hell-Date test set, which is the one trained first on PLL then on Hell-Date.

We can interpret the results in terms of MAE presented in Table~\ref{tab:results_pre_training_HD} by focusing on the case where the ETW width equals the MAE. With an approximate MAE of 55 years, the relative ETW spans approximately 110 years. Under this time tolerance, the network correctly dates nearly $60\%$ of the test documents. Notably, Figure~\ref{fig:error_wind} shows that when the ETW width decreases to 50 years ($\alpha=0.5$), the accuracy remains relatively stable at around $50\%$. This suggests that despite the average error, the predictions are accurate for a significant proportion of documents, so we can expect the network to give accurate estimations of the century of production of handwriting.

\begin{figure}[ht]
\centering
\includegraphics[width=1\textwidth]
{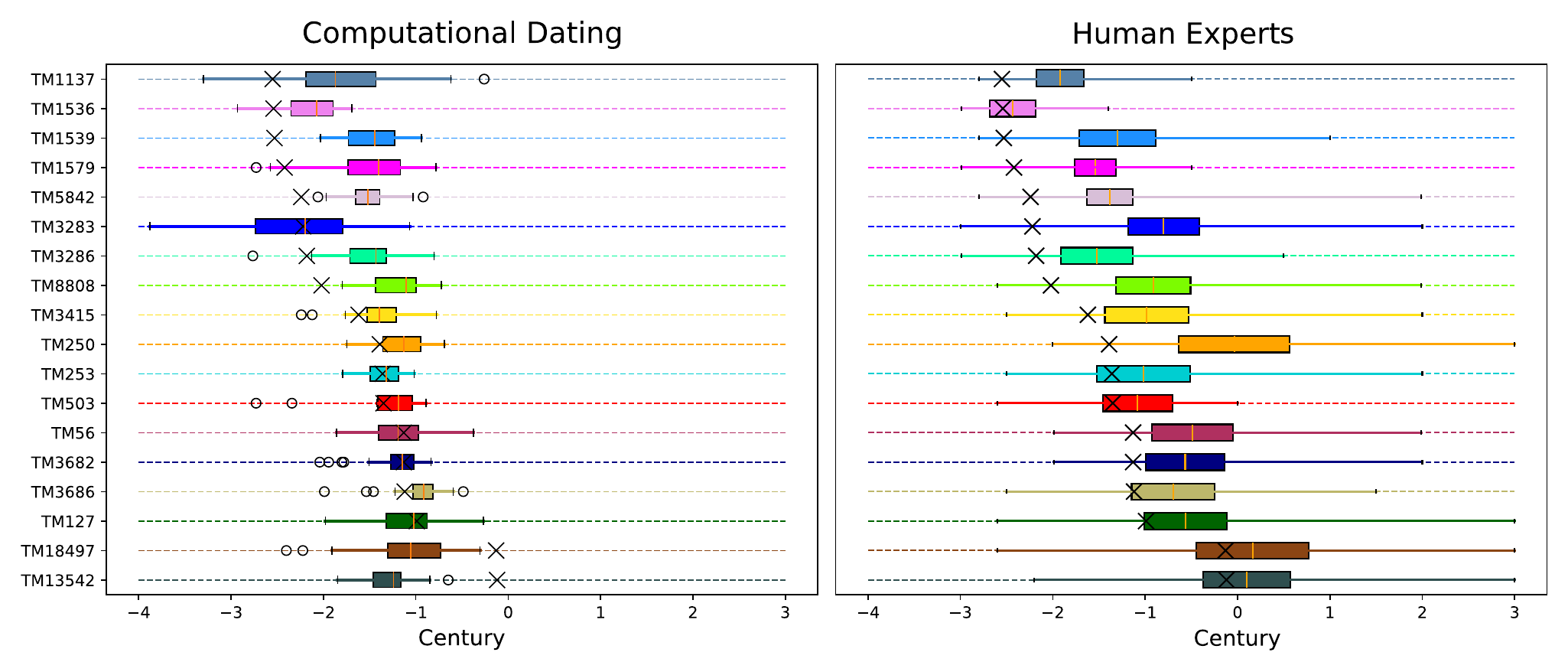}
\caption{Datings by fCNN (on the left) and the experts (on the right) a box plots per document. The ground truth (actual date of writing) is indicated with an X.}
\label{fig:error_dist_HD}
\end{figure}

Looking further at Figure~\ref{fig:error_wind}, we observe that the network achieves an accuracy of 1 only with an error window width equal to two and a half centuries ($\alpha=2.5$). Given the vast breadth of this time interval, it is worth analysing further the error distribution of the network predictions for single lines. The left plot in Figure~\ref{fig:error_dist_HD} shows the time-axis distribution of the predictions for individual lines of each document in the test set. Documents are chronologically ordered, starting on the top with the most ancient. While some documents exhibit less precise dating, over half of the test cases display very accurate average predictions. Additionally, there appears to be a trend towards greater precision when dating in the second half of the 2\textsuperscript{nd} c. BCE.

\subsection{Human-AI Comparison}
To compare the outcomes achieved by both the expert scholars and the computational model, we use the MAE values obtained by the best-performing model and the ones of humans without excluding empty answers. To facilitate comparison, we report in the right plot of Figure~\ref{fig:error_dist_HD} the error distribution made by expert scholars for each document. Table~\ref{tab_TM_error} sums up the comparison, giving the MAE and standard deviation $\sigma$ of the predictions for each document, as committed by both the computational model and the human experts. With the exception of TM~1137 and 1536, the two most ancient documents in the test set, the computational approach yields better results in terms of MAE while keeping a lower standard deviation.

\begin{table}[ht]
\centering
\caption{Results (MAE and standard deviation $\sigma$) of the prediction for each document for the AI model, the mean human experts, and the best-performing human expert.}
\label{tab_TM_error}
\begin{tabular}{l|cc|cc|cc}
& \multicolumn{2}{c}{} & \multicolumn{2}{c}{}  & \multicolumn{2}{c}{}\\
& \multicolumn{2}{c}{AI-Model} & \multicolumn{2}{c}{Human Experts}  & \multicolumn{2}{c}{Human Expert H2}\\
\hline 
\textbf{TM} & \textbf{MAE} & \textbf{$\sigma$} & \textbf{MAE} & \textbf{$\sigma$} &\textbf{MAE} & \textbf{$\sigma$}\\
\hline\hline
TM1137  & 0.78 & 0.57 & 0.70 & 0.53 & 0.15 & 0.19 \\
TM1536  & 0.47 & 0.34 & 0.14 & 0.34 & 0.21 & 0.47 \\
TM1539  & 1.06 & 0.32 & 1.29 & 0.98 & 0.58 & 0.42 \\
TM1579  & 0.94 & 0.39 & 1.09 & 0.61 & 0.24 & 0.25 \\
TM5842  & 0.72 & 0.28 & 1.07 & 1.42 & 0.10 & 0.17 \\
TM3283  & 0.52 & 0.65 & 1.59 & 1.43 & 0.09 & 0.24 \\
TM3286  & 0.71 & 0.39 & 0.81 & 0.82 & 0.20 & 0.25 \\
TM8808  & 0.84 & 0.28 & 1.27 & 1.28 & 0.29 & 0.50 \\
TM3415  & 0.32 & 0.31 & 0.74 & 0.91 & 0.09 & 0.36 \\
TM250   & 0.30 & 0.29 & 1.40 & 1.29 & 0.58 & 0.87 \\
TM253   & 0.17 & 0.20 & 0.75 & 1.09 & 0.28 & 0.40 \\
TM503   & 0.42 & 0.62 & 0.73 & 0.67 & 0.27 & 0.53 \\
TM3682  & 0.19 & 0.29 & 1.03 & 1.10 & 0.39 & 0.43 \\
TM56    & 0.25 & 0.31 & 0.91 & 1.20 & 0.37 & 0.60 \\
TM3686  & 0.23 & 0.21 & 0.77 & 0.96 & 0.35 & 0.54 \\
TM127   & 0.29 & 0.37 & 1.05 & 1.40 & 0.61 & 0.67 \\
TM18497 & 0.92 & 0.43 & 1.39 & 1.74 & 1.17 & 0.75 \\
TM13542 & 1.16 & 0.27 & 1.43 & 1.33 & 1.41 & 0.41 \\
\hline
\end{tabular}
\end{table}

\begin{figure}[ht]
\centering
\includegraphics[width=0.7\textwidth]{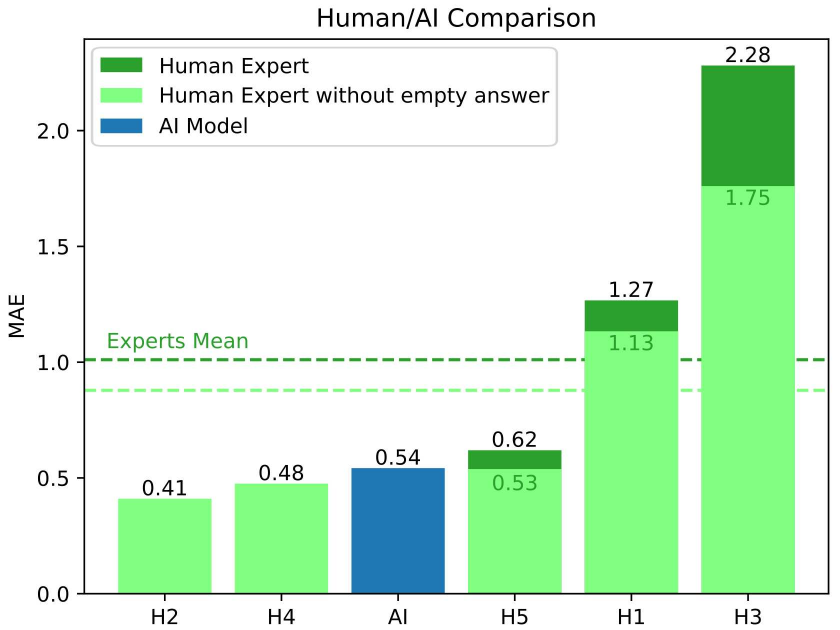}
\caption{Comparison between human experts and AI Model performance.}
\label{fig:AIH_comparison}
\end{figure}

However, the results of the individual experts varied heavily. Therefore, we produced Figure~\ref{fig:AIH_comparison} summarising all results in terms of MAE. For the sake of thoroughness, we included both MAE values with and without empty answers. It is noticeable that while the computational model demonstrates superior performance compared to the average human result, only two experts achieved surpassing the AI. The confidence of these experts is further visible by the fact that they did not leave any empty answer.

\begin{figure}[ht]
\centering
      \includegraphics[width=\textwidth]{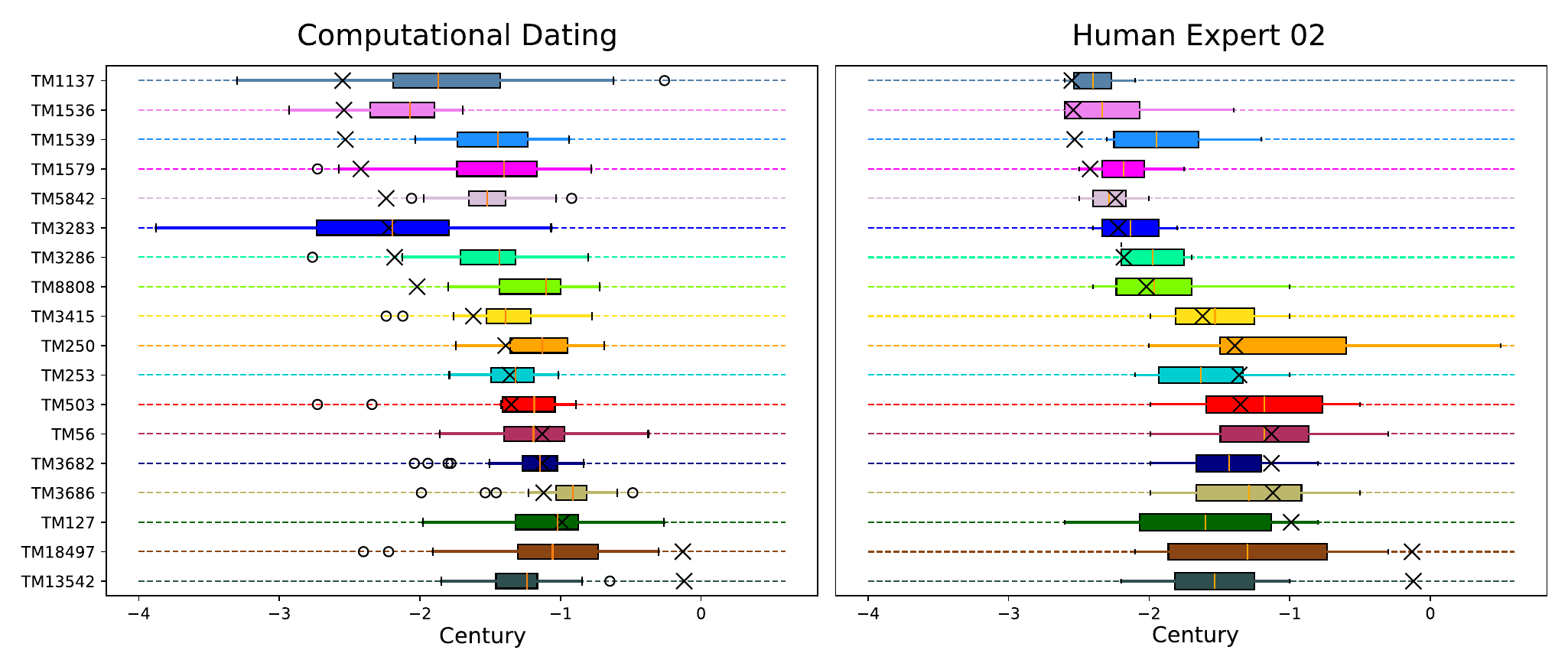}
\caption{Comparison between computational model performance (left) and best-performing Human Expert (right).}
\label{fig:AIH_comparison_best}
\end{figure}

At this point, it is relevant to directly compare the results of the computational model with those from the best expert (H2). Figure~\ref{fig:AIH_comparison_best} shows the error distribution for each. In general, the expert tends to give larger estimations than the AI, as can also be seen for the $\sigma$ and MAE values reported in Table~\ref{tab_TM_error}. Therefore, the ground truth falls in the two best quartiles in twelve cases compared to six cases for the computational model. This fact suggests that even the best-performing human tends to give a larger time span in order to increase the probability that the ground truth falls in the given range. Moreover, the human has a less marked tendency to date papyri in the 2\textsuperscript{nd} c. BCE than the computational approach.

To look more deeply at the differences, we had a look at some of the papyri in the test set. The best-performing human dates better documents from the 3\textsuperscript{rd} c. BCE. Compared to them, fCNNr poorly performed in dating these early documents and tended to attribute them to the following century. Although not all humans outperformed the AI model in dating documents from the 3\textsuperscript{rd} c. BCE, it is clear from the human mean that experts usually correctly date one document: TM~1536. This papyrus is written in a style that is well studied in papyrological scientific literature, the so-called Alexandrian chancery style \cite[26--31]{cavallo_2008}, a fact that could explain the good results of human experts. The machine outperforms the best human for documents dated in the late 2\textsuperscript{nd} c. BCE, TM~503, 253 and 250. These three papyri come from an ensemble of papyri penned by the same two writers, a man called Dryton and his son Esthladas (see \cite{vandorpe_2002}). Papyri from the same ensemble are present in the training set. The high precision of the model in dating these three documents suggests that scribal identification may play a role in the dating process. In the case of TM~253, the papyrus is today divided into two fragments, one preserved in London, the other in Heidelberg. The image of the London fragment was in the training set, and the image of the Heidelberg one was in the test set. One may suggest that, despite the difference in preservation and digitisation of the two fragments (for instance, in the colour scale and brightness of the two images), the network managed to identify identical scripts which led to a relatively precise prediction. Finally, for the two most recent documents, TM~13542 and 18497, both the fCNNr network and the best-performing human wrongly located them in the middle of the possible time span. On the contrary, the average answer of human experts correctly dated these documents. However, we noticed high uncertainty of the human predictions and strong disagreement among the experts concerning these two documents. 
Among our respondents, some had only expertise in Hellenistic papyri and some in Hellenistic and Roman papyri. As these very late documents' writing is close to Roman cursive scripts, the experts with knowledge of Roman scripts may have been facilitated in dating those pieces compared to those with exclusively Hellenistic expertise.

%% file: texs/08_Conclusion.tex
In this study, we introduced a novel, precisely dated dataset named Hell-Date. We used fCNN to predict dates and observed only little improvement using transfer learning. To evaluate the performance of dating models in the context of historical documents, we propose a framework that incorporates the inherent imprecision of the palaeographic dating method. It relates the accuracy of the prediction with the size variation of the Error Time Window and integrates for each document the variability of the prediction, providing an in-depth view of the predictive capabilities of the model. The comparison with human results shows that current models are already able to give results comparable to those of experts. Future works will aim on the one hand at covering a larger time period with a more numerous dataset and on the other hand at better defining handwriting style similarities to provide new interpretations of their variety.

%% file: main.bbl
\begin{thebibliography}{10}
\providecommand{\url}[1]{\texttt{#1}}
\providecommand{\urlprefix}{URL }
\providecommand{\doi}[1]{https://doi.org/#1}

\bibitem{adam2018kertas}
Adam, K., Baig, A., Al-Maadeed, S., Bouridane, A., El-Menshawy, S.: Kertas: Dataset for automatic dating of ancient arabic manuscripts. International Journal on Document Analysis and Recognition (IJDAR)  \textbf{21},  283--290 (2018)

\bibitem{bagnall_2009}
Bagnall, R.S.: Practical help: chronology, geography, measures, currency, names, prosopography, and technical vocabulary. In: Bagnall, R.S. (ed.) The {Oxford} Handbook of Papyrology, pp. 179--196. Oxford University Press, Oxford, New York (2009)

\bibitem{baledent-etal-2020-dating}
Baledent, A., Hiebel, N., Lejeune, G.: Dating ancient texts: an approach for noisy {F}rench documents. In: Sprugnoli, R., Passarotti, M. (eds.) Proceedings of LT4HALA 2020 - 1st Workshop on Language Technologies for Historical and Ancient Languages. pp. 17--21. European Language Resources Association (ELRA), Marseille, France (May 2020), \url{https://aclanthology.org/2020.lt4hala-1.3}

\bibitem{Bennet_2011}
Bennett, C.: Alexandria and the moon: an investigation into the lunar {M}acedonian calendar of {P}tolemaic {E}gypt. Peeters, Leuven (2011)

\bibitem{cavallo_2008}
Cavallo, G.: La scrittura greca e latina dei papiri: una introduzione. F. Serra, Pisa, Rome (2008)

\bibitem{cavallo_2009}
Cavallo, G.: {Greek} and {Latin} writing in the papyri. In: Bagnall, R.S. (ed.) The {Oxford} Handbook of Papyrology, pp. 101--148. Oxford University Press, Oxford, New York (2009)

\bibitem{choat_2019}
Choat, M.: Dating papyri: Familiarity, instinct and guesswork. Journal for the Study of the New Testament  \textbf{42}(1),  58--83 (2019). \doi{10.1177/0142064X19855580}

\bibitem{Cloppet2017}
Cloppet, F., Eglin, V., Helias-Baron, M., Kieu, C., Vincent, N., Stutzmann, D.: Icdar2017 competition on the classification of medieval handwritings in latin script. In: 2017 14th IAPR International Conference on Document Analysis and Recognition (ICDAR). vol.~01, pp. 1371--1376 (2017). \doi{10.1109/ICDAR.2017.224}

\bibitem{trismegistos}
Depauw, M., Gheldof, T.: Trismegistos: An interdisciplinary platform for ancient world texts and related information. In: Bolikowski, {\L}., Casarosa, V., Goodale, P., Houssos, N., Manghi, P., Schirrwagen, J. (eds.) Theory and Practice of Digital Libraries -- TPDL 2013 Selected Workshops. pp. 40--52. Springer International Publishing, Cham (2014)

\bibitem{DHALI2020413}
Dhali, M.A., Jansen, C.N., {de Wit}, J.W., Schomaker, L.: Feature-extraction methods for historical manuscript dating based on writing style development. Pattern Recognition Letters  \textbf{131},  413--420 (2020). \doi{10.1016/j.patrec.2020.01.027}, \url{https://www.sciencedirect.com/science/article/pii/S0167865520300386}

\bibitem{hamid2019deep}
Hamid, A., Bibi, M., Moetesum, M., Siddiqi, I.: Deep learning based approach for historical manuscript dating. In: 2019 International Conference on Document Analysis and Recognition (ICDAR). pp. 967--972. IEEE (2019)

\bibitem{harrauer_2010}
Harrauer, H.: Handbuch der griechischen {Paläographie}. A. Hiersemann, Stuttgart (2010)

\bibitem{li2015publication}
Li, Y., Genzel, D., Fujii, Y., Popat, A.C.: Publication date estimation for printed historical documents using convolutional neural networks. In: Proceedings of the 3rd international workshop on historical document imaging and processing. pp. 99--106 (2015)

\bibitem{docextractor}
Monnier, T., Aubry, M.: docextractor: an off-the-shelf historical document element extraction. In: 2020 17th International Conference on Frontiers in Handwriting Recognition (ICFHR). pp. 91--96 (2020). \doi{10.1109/ICFHR2020.2020.00027}

\bibitem{nongbri_2019}
Nongbri, B.: Palaeographic analysis of codices from the early christian period: A point of method. Journal for the Study of the New Testament  \textbf{42}(1),  84--97 (2019). \doi{10.1177/0142064X19855582}

\bibitem{orsini_clarysse_2012}
Orsini, P., Clarysse, W.: Early new testament manuscripts and their dates: a critique of theological palaeography. Ephemerides Theologicae Lovanienses  \textbf{88},  443--474 (2012). \doi{10.2143/ETL.88.4.2957937}

\bibitem{paparrigopoulou2023greek}
Paparrigopoulou, A., Kougia, V., Konstantinidou, M., Pavlopoulos, J.: Greek literary papyri dating benchmark. In: Coustaty, M., Forn{\'e}s, A. (eds.) Document Analysis and Recognition -- ICDAR 2023 Workshops. pp. 296--306. Springer Nature Switzerland, Cham (2023)

\bibitem{Paulissen_Vandorpe_2019}
Paulissen, J., Vandorpe, K.: Dating early {P}tolemaic salt tax receipts: the {E}gyptian tax year. Zeitschrift für Papyrologie und Epigraphik  \textbf{211},  145--161 (2019), \url{https://www.jstor.org/stable/48632501}

\bibitem{pavlopoulos-etal-2023-dating}
Pavlopoulos, J., Konstantinidou, M., Marthot-Santaniello, I., Essler, H., Paparigopoulou, A.: Dating {G}reek papyri with text regression. In: Rogers, A., Boyd-Graber, J., Okazaki, N. (eds.) Proceedings of the 61st Annual Meeting of the Association for Computational Linguistics (Volume 1: Long Papers). pp. 10001--10013. Association for Computational Linguistics, Toronto, Canada (Jul 2023). \doi{10.18653/v1/2023.acl-long.556}, \url{https://aclanthology.org/2023.acl-long.556}

\bibitem{pavlopoulos2023explaining}
Pavlopoulos, J., Konstantinidou, M., Vardakas, G., Marthot-Santaniello, I., Perdiki, E., Koutsianos, D., Likas, A., Essler, H.: Explaining the chronological attribution of greek papyri images. In: International Conference on Discovery Science. pp. 401--415. Springer (2023). \doi{10.1007/978-3-031-45275-8\_27}

\bibitem{Seuret2021}
Seuret, M., Nicolaou, A., Rodr{\'i}guez-Salas, D., Weichselbaumer, N., Stutzmann, D., Mayr, M., Maier, A., Christlein, V.: Icdar 2021 competition on historical document classification. In: Llad{\'o}s, J., Lopresti, D., Uchida, S. (eds.) Document Analysis and Recognition -- ICDAR 2021. pp. 618--634. Springer International Publishing, Cham (2021)

\bibitem{vandorpe_2002}
Vandorpe, K.: The bilingual family archive of {Dryton}, his wife {Apollonia} and their daughter {Senmouthis} ({P}.{Dryton}). Koninklijke Vlaamse Academie van België voor Wetenschappen en Kunsten, Brussels (2002)

\bibitem{Wahlberg2016hist}
Wahlberg, F., Wilkinson, T., Brun, A.: Historical manuscript production date estimation using deep convolutional neural networks. In: 2016 15th International Conference on Frontiers in Handwriting Recognition (ICFHR). pp. 205--210 (2016). \doi{10.1109/ICFHR.2016.0048}

\end{thebibliography}
